# Quantitative In Vivo Cherenkov Luminescence Imaging and Dosimetry of $^{86}$Y-NM600

*Campbell D. Haasch[1], Malick Bio Idrissou, Sydney Jupitz[1], Aubrey Parks[1], Reinier Hernandez[1,2,3,4], Brian W. Pogue[1,5], Bryan P. Bednarz[1,2,3,4]

[1]*Department of Medical Physics, University of Wisconsin School of Medicine and Public Health, University of Wisconsin–Madison, Madison, WI, United States*

[2]*Department of Human Oncology, University of Wisconsin School of Medicine and Public Health, University of Wisconsin–Madison, Madison, WI, United States*

[3]*Department of Radiology, University of Wisconsin School of Medicine and Public Health, University of Wisconsin–Madison, Madison, WI, United States*

[4]*Carbone Cancer Center, University of Wisconsin–Madison, Madison, WI, United States*

[5]*Thayer School of Engineering at Dartmouth, Hanover, New Hampshire 03755, USA*

**ORCID IDs:** Campbell Haasch (0009-0002-7128-0129), Malick Bio Idrissou (0009-0001-0437-6668), Sydney Jupitz (0000-0001-9104-8920), Aubrey Parks (0009-0000-9382-0937), Reinier Hernandez (0000-0002-0729-2179), Brian Pogue (0000-0002-9887-670X), Bryan Bednarz (0000-0002-7467-9816)

*Corresponding author

Address: 1111 Highland Avenue, 7109, Madison, WI 53705

Email: chaasch@wisc.edu

# ABSTRACT

**Purpose**

The rapid expansion of radiopharmaceutical therapy (RPT) development demands scalable preclinical dosimetry methods. While PET and SPECT remain the gold standards, their low throughput and high cost limit large-cohort studies. Cherenkov luminescence imaging (CLI) offers a high-throughput alternative but suffers from depth-dependent attenuation and photon scatter that compromise quantitative accuracy. This work develops and validates a quantitative CLI methodology incorporating attenuation and scatter corrections to enable accurate preclinical dosimetry.

**Methods**

Depth-dependent attenuation was characterized using a tissue-mimicking phantom to derive calibration coefficients. Photon scatter was modeled using GEANT4-generated Cherenkov spread functions (CSFs), applied in a depth-weighted iterative Richardson–Lucy deconvolution/reconvolution framework. The method was evaluated in NU/NU mice (n=4) bearing MC38 tumors after injection of $^{86}$Y-NM600, an isotope suitable for both PET and CLI. Liver and tumor activities were quantified at four timepoints using PET and the proposed CLI method. Voxelized Monte Carlo dosimetry was performed for both modalities.

**Results**

CLI–PET activity quantification yielded mean errors of 15.4% (liver) and 10.3% (tumor) over the first three timepoints. Tumor absorbed doses from CLI-derived synthetic PET images (3.4 ± 0.3 Gy/MBq) were statistically indistinguishable from PET-based estimates (3.2 ± 0.2 Gy/MBq, p=0.31). Discrepancies increased at late timepoints due to low activity and background auto-luminescence.

**Conclusions**

With appropriate depth-dependent attenuation calibration and Monte Carlo–derived scatter correction, CLI can provide quantitative biodistribution and dosimetry estimates comparable to PET. This approach enables high-throughput, low-cost in vivo dosimetry, expanding the feasibility of large-scale preclinical RPT studies and supporting translational radiopharmaceutical development.

# INTRODUCTION

The recent FDA approvals of [$^{177}$Lu]-DOTA-TATE (Lutathera) and [$^{177}$Lu]-PSMA-617, (Pluvicto) have marked the beginning of a revitalized era of radiopharmaceutical therapy (RPT) development [1]. The pace of radiopharmaceutical development has increased not only for agents targeting PSMA and SSTR, but also for a variety of novel tumor-associated targets, including FAP, CA-IX, Nectin-4, GRPr, members of the integrin family, and many others [2-3]. Furthermore, emerging therapeutic strategies such as pre-targeting approaches and targeted alpha therapy are continuing to expand the paradigms of RPT applications. The rapid expansion of the RPT pipeline necessitates advancement in preclinical drug evaluation methodologies. Specifically, accurate determination of absorbed dose delivered by ionizing radiation in animal models is essential for assessing both therapeutic efficacy and normal tissue toxicity. Absorbed dose measurements are not only applicable to understanding biological effects in preclinical tumor models but also for informing dose ranges for translational Phase 1 escalation trials [4].

Preclinical dosimetry is traditionally performed using serial PET or SPECT imaging to quantify the time-dependent, three-dimensional biodistribution of radiotracers. Absorbed dose calculations are then conducted using organ-level S-values, point dose kernels, or Monte Carlo (MC) simulations applied to the imaging-derived activity distribution [5]. While PET and SPECT are well-established and validated modalities for absorbed dose determination, they present significant throughput limitations, typically accommodating a maximum of four mice per 30-60 minute imaging session. Additionally, pure beta emitters and alpha emitters are not amenable to preclinical imaging, necessitating the use of surrogate imaging isotopes—a practice that may introduce systematic errors between imaging and therapeutic compounds.

Optical imaging represents an alternative, albeit less extensively explored, approach to conventional nuclear imaging modalities for dosimetric applications [6]. Beta emissions in dielectric media exceeding the phase velocity of light – approximately 240 keV in tissue generate Cherenkov radiation, characterized by broad-spectrum emission spanning ultraviolet to far-infrared wavelengths with intensity following a $1/\lambda^2$ dependence [7]. While the $1/\lambda^2$ dependence causes blue weighted emission, it is primarily read-near infrared (NIR) light that escapes from tissue due to the presence of an optical window in this regime [8]. Investigations by several research groups demonstrated the feasibility of Cherenkov luminescence imaging (CLI) in murine models using conventional bioluminescence imaging systems equipped with thermoelectrically cooled charge-coupled device (CCD) cameras [9-12]. CLI is a high-throughput modality capable of imaging 5-10 mice in a short time frame and has the potential to offer preclinical dosimetry at scales not achievable using conventional nuclear imaging.

However, the use of CLI in dosimetry remains underdeveloped since in vivo activity quantification is challenging due to the significant attenuation and scattering of optical photons in tissue [13]. These challenges must be accounted for to ensure proper calibration of each measurement and prevent signal cross-contamination from adjacent sources. While attenuation coefficient-based corrections have been implemented in previous works, no established methodology exists to address photon diffusion effects, which result in depth-dependent blurring of optical signals and compromise quantitative accuracy [14].

To address these limitations, we have developed a quantitative model that overcomes the challenges inherent in Cherenkov luminescence imaging. We addressed the attenuation of Cherenkov photons through phantom-based calibrations coupled with a priori anatomical imaging for Region-of-Interest(ROI)-specific corrections. Additionally, we addressed photon scattering effects through iterative deconvolution algorithms employing GEANT4-generated Cherenkov point spread functions. The quantitative accuracy of our model was validated using a murine model administered $^{86}$Y-labeled alkyl phosphocholine NM600. $^{86}$Y represents an optimal isotope for comparative evaluation of PET and CLI methodologies due to its high-energy positron emission, which generates substantial Cherenkov radiance while maintaining compatibility with PET imaging. Activity concentrations in primary uptake regions, specifically liver and tumor tissues, were quantified using both conventional PET imaging and the developed CLI model. Finally, tumor-absorbed doses were calculated using a MC methodology to demonstrate that CLI provides dosimetric estimates with comparable accuracy to the established PET gold standard.

## MATERIALS AND METHODS

### Attenuation Modeling

The radiance of Cherenkov photons generated from activity at different depths is a complex problem due to its dependence on the coupling of both electron and optical photon transport. The Beer-Lambert law is typically used to express the effects of absorption and scattering on a monoenergetic photon intensity as follows[14]:

$$I(\lambda) = I_0(\lambda) \exp(-\mu_{eff}(\lambda) * d) \tag{1}$$

Where $\mu_{eff}$ is the wavelength-dependent effective attenuation coefficient as a function of the absorption coefficient, $\mu_a$, and the reduced scattering coefficient, $\mu_s'$:

$$\mu_{eff} = \sqrt{3\mu_a(\mu_s' + \mu_a)} \tag{2}$$

This model is difficult to apply to Cherenkov attenuation as Cherenkov radiance is a broad multispectral emission, and production is influenced by transport of generating beta particles in addition to optical photons While a narrow bandpass filter may be used to approximate a monoenergetic emission spectra and narrow the attenuation coefficient to a scalar value, this reduces the total signal acquired from an already low radiance source.

The number of Cherenkov photons is dependent on the pathlength traveled by a beta particle as defined by the following equation:

$$N(\lambda, \beta, n) = 2\pi\alpha \frac{1}{\lambda}\left(1 - \frac{1}{\beta^2 n^2}\right) \qquad (3)$$

Activity at deeper locations in tissue will see greater optical photon attenuation, however, there will also be a greater number of Cherenkov photons generated until the depth at which most electrons can traverse their full path length with energies > 240 keV, as shown in **Figure 1.**

Due to these challenges we approach the depth-based calibration of the Cherenkov radiance using a model agnostic approach, instead taking phantom measurements to create a calibration curve not strictly defined by previous theory.

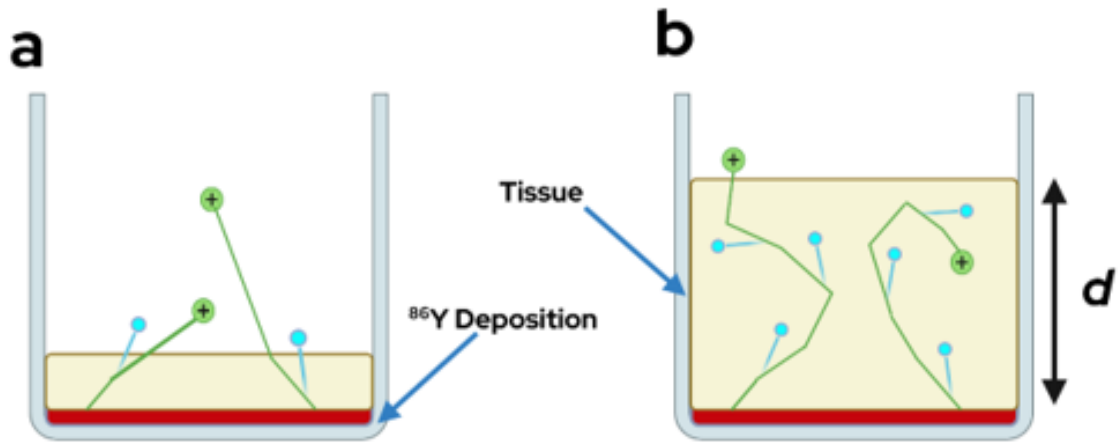

**Figure 1.** Well diagrams of $^{86}$Y deposited below a depth, d, of a tissue equivalent mixture. Activity at a shallower depth in tissue (A) will produce less Cherenkov photons due to a shorter path length in the media but experience less attenuation while activity at a deeper depth (B) will produce maximal Cherenkov but experience greater attenuation.

## Phantom Study

We developed an imaging phantom using a 96-well FLUOTRAC well plate (Greinier Bio-One, Kremsmünster, Austria). Five varying quantities of $^{86}$YCl$_3$ were plated in triplicate (0, 0.185, 0.370, 0.925, 1.850 MBq) in 100 µL of 140 Proof EtOH in across the well plate. After plating, the activity was left to dry and adhere to the bottom of the well plate for one hour on a heating block. To mimic the optical properties of biological tissue, we used an optical mixture consisting of 1% soybean intralipid (Sigma Aldrich, St. Louis, MI), 1% Bovine Whole Blood (Lampire, Pippersville, PA) and 98% deionized water. The tissue mimicking mixture has previously been used as a biological tissue phantom in external beam radiotherapy Cherenkov studies [15].

CLI of the well plate phantom was performed using an IVIS Spectrum Optical System (Revvity, Waltham, MA). We performed CLI imaging of the well plate filled with varying depths of optical mixture (1.66, 3.31, 4.97, 6.62, 8.28, 9.93 mm). Each well plate image was captured over 300 seconds (F number: 1, binning factor: 8, FOV: 13.3 cm) with the emission filter open. The radiance in each well was measured using the software Living Image (Revvity, Waltham, Ma). Decay corrections were applied between the time of plating and each image acquisition.

## GEANT4 Generation of Cherenkov Spread Functions

Optical simulations in the MC code Geant4 were performed to model the transport of $^{86}$Y-generated Cherenkov photons and create Cherenkov Spread Functions (CSF) to characterize the diffusion of Cherenkov light from a point source at depth in tissue[16]. Geant4 and its associated derived codes are appropriate for the simulation of Cherenkov light production in tissue due to their implementation of both electronic and optical transport.

Geant4 does not implement material absorption and scattering coefficients or refractive indices in the low-energy optical photon range. Instead, the user must express each optical property as a function of photon energy for each material used in the simulation.

The refractive index (n) influences both the energy threshold for Cherenkov radiation generation and the reflection of Cherenkov photons at tissue–air interfaces. Although the refractive index is wavelength-dependent, it varies by only a few percent in skin across the visible to near-infrared spectrum [17]. Previous studies across various species and tissue layers have reported refractive index values ranging from 1.34 to 1.57. For our model, we defined the refractive index of mouse skin as 1.4 [17, 18]. The user must additionally define the absorption mean free path (MFP) ($\frac{1}{\mu_a}$), the Rayleigh scattering MFP ($\frac{1}{\mu_{s,Ray}}$), and the Mie scattering MFP ($\frac{1}{\mu_{s,Mie}}$).

The absorption and scattering mean free paths are parameterized and tissue dependent coefficients provided in Jacques 2013 [13]. The wavelength dependent absorption coefficient is defined by the concentration of a tissue's primary chromophores: Oxygenated Blood (BS), deoxygenated blood (B(1-S)), water (W), Bilirubin ($C_{Bili}$), Beta-Carotene ($C_{\beta C}$), and Melanin (M) and the wavelength dependent absorption spectra of the chromophores. For our use cases, Bilirubin, Beta-Carotene, and Melanin were assumed to be negligible due to their low concentration in nude mice skin. The parameterized equation for the absorption coefficient is given as the following:

$$\mu_a = BS\mu_{a,oxy}(\lambda) + B(1-S)\mu_{a,deoxy}(\lambda) + W\mu_{a,water}(\lambda) + F\mu_{a,fat}(\lambda) \tag{4}$$

The wavelength dependent absorption coefficients were obtained from [19, 20] (**Figure 2a**).

The scattering theory implemented in our MC simulation is based on the division of scattering into Rayleigh (isotropic and attributed to small scatterers) and Mie (anisotropic and attributed to scatters on the order of the wavelength of scattered light) scattering processes. The Rayleigh scattering coefficient is parameterized in Eq. (5) by the tissue dependent Rayleigh scattered fraction ($f_{Ray}$), and scaling factor ($a'$) defined as the scattering coefficient at 500 nm. The coefficient is scaled by the -4th power familiar to the Rayleigh scattering process.

$$\mu'_{s,Ray} = a'\left(f_{Ray}\frac{\lambda}{500\ nm}\right)^{-4} \tag{5}$$

The Mie scattering coefficient is parameterized similarly in Eq.(6) with the addition of a tissue dependent scattering power $-b_{mie}$.

$$\mu'_{s,Mie} = a'\left((1-f_{Ray})\frac{\lambda}{500\text{nm}}\right)^{-b_{Mie}} \quad (6)$$

Additionally, the anisotropy of Mie scattering is approximated in GEANT4 using a double Henyey-Greenstein function with an anisotropy parameter for both forward and backward scattering, $g_F$ and $g_B$. Additionally, the ratio of forward to backward scatter, $Mie_{FB}$, must be defined. Mie scattering parameters were assumed to be water equivalent. The Geant4 model used to generate the CSFs was given the optical coefficients of skin listed in **Table 1**. The computed absorption and total scattering coefficients of the Geant4 skin phantom are displayed in **Figure 2b**.

**Table 1. Optical parameters used to characterize scattering and absorption of Cherenkov light in the range 400-1000 nm in the skin for Geant4 optical simulations. Parameters are taken from reported literature values [13].**

| Optical Parameter | Value |
|---|---|
| B | 0.34 |
| S | 98.5 |
| W | 21.4 |
| F | 27.7 |
| $a'(cm^{-1})$ | 48.0 |
| $f_{Ray}$ | 0.409 |
| $b_{mie}$ | 0.702 |
| $g_F$ | 0.91 |
| $g_B$ | 0.87 |
| $Mie_{FB}$ | 0.85 |

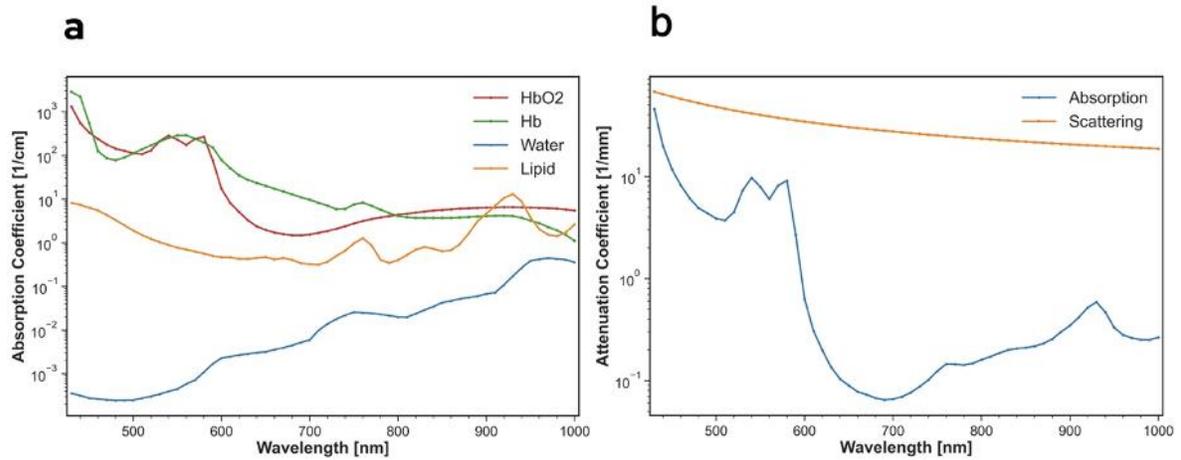

**Figure 2.** (A) Absorption coefficients of tissue components over visible to near-infrared range (430-1000 nm). (B) Computed total scattering and absorption coefficients for use in Geant4 simulation of Cherenkov diffusion.

The geometry of the CSF simulation consisted of a 2 cm x 2 cm x 2 cm tissue phantom in air (**Figure 3**). We defined scoring volume with a thickness of 0.1 cm on the top surface of the tissue phantom. Each Cherenkov photon that crossed the boundary of the scoring volume had its crossing position recorded and was subsequently killed. The flux across the scoring volume was recorded in list mode so that arbitrary binning could be applied post hoc to transfer the CSF to any alternative pixel size. CSFs with source depths of 2-10mm were generated.

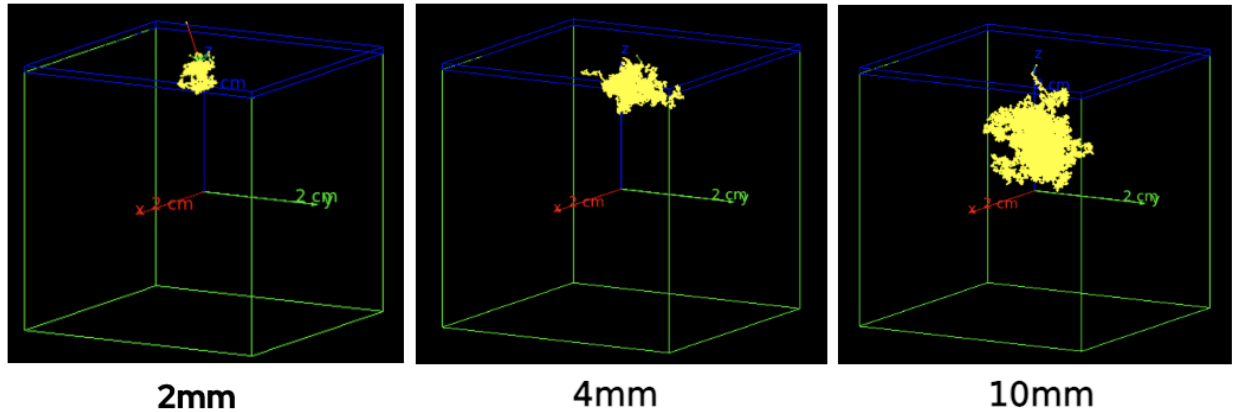

**Figure 3.** Representative simulations of $^{86}$Y positron emission at varying depths in tissue phantom. The tissue phantom is depicted in green while the scoring volume is depicted in blue. Cherenkov photons crossings into scoring volume are recorded and crossing photons are killed. As depth increase, photons diffuse a greater distance before reaching the scoring volume.

**Animal Study**

The animal study was performed under the approval of the University of Wisconsin Institutional Animal Care and Use Committee. NU/NU mice (n=4) bearing MC38 xenografts were injected with 9.25 MBq $^{86}$Y-NM600 via the tail vein. Previous works have described the synthesis and radiolabeling of $^{86}$Y-NM600 [21-22]. Serial PET scans (3.5, 29, 52, 97 hours post injection) were acquired of each subject using an Inveon small-animal μPET/CT (Siemens Medical Solutions, Erlangen, Germany). Mice were anesthetized with isoflurane (2.5% induction reduced to 1.5% for maintenance). Scans of each mouse were stopped after 80 million coincidence events were recorded (energy window 350-650 keV; time window 3.432 ns) with CT scans taken directly after (80 kVp, 120 projections). PET images were reconstructed using a 3D OSEM algorithm. Co-registered CTs were used to calculate attenuation corrections in the PET scan reconstruction as well as being later used as CLI anatomical reference and as tissue density reference in dose calculations. The mice were transported on a stable black plastic sheet from the Inveon bed to an IVIS Spectrum (Revvity, Waltham, MA) optical imaging platform while still under the residual effects of anesthesia. With this approach the mice remained in the same position during transport ensuring the CT and CLI images could be accurately registered. While on the optical imaging platform, the mice were once again placed under 1.5% isoflurane. CLI images of each mouse were captured in both prone and supine views (Exposure time: 300 s, FOV: 13.3 cm, F#: 1, binning factor: 8). Liver and tumor ROIs were contoured on each CT. Contoured ROIs translated to the registered PET image and used to estimate ROI uptake at serial timepoints.

## CLI Activity Quantification

### 1. Image Registration

Our proposed activity quantification methodology uses the previously described photon propagation modeling in combination with anatomical micro-CT data to quantify Cherenkov luminescence images. Reducing the CT data from three dimensions to two was necessary to enable registration with the planar CLI images. A mask of each mouse at each timepoint was created by applying a threshold of -700 HU to each CT. The mask was then reduced to the coronal view via a maximum intensity projection (MIP) along the sagittal axis. Similar masks of the prone and supine CLI images were created by placing a threshold on the gray value of each mouse white light photograph. CLI masks were cleaned post-thresholding using manual methods to ensure masking was accurate.

The CLI images were scaled to the resolution of the CT images (CLI Resolution: 0.554 mm, CT Resolution 0.204 mm). Rescaling the CLI images has no effect on total radiance as each CLI image is reported in area normalized units (photons/s/cm$^2$/sr). Two-dimensional similarity registration (Scale, translation, and rotation) was performed in python using SimpleITK over 100 iterations. Manual registration was performed after automated registration to ensure CLI images entirely aligned with contoured volumes (**Figure 4b**). Prone CLI registration was direct as the mice were transported between the PET/CT and IVIS bed in the same prone position. For the supine view, however, there was no corresponding CT image. The prone CT was rotated 180º about the longitudinal axis when creating the supine mask for registration. Due to the synthetic CT supine positions being unmatched with the supine CLI, a greater degree of manual registration was required to ensure the liver ROI aligned with the liver in the CT image.

### 2. *ROI Depth Map Generation*

To encode anatomical depth from the registered CT, we generated planar ROI projections where each pixel stores the average depth (mm) of the underlying voxels measured from the mouse surface (**Figure 4c**). The voxel depth was calculated using the external contour of the CT mask and the labeled liver and tumor volumes. The voxel-wise distance between each voxel in the ROI and the surface directly above it was calculated and multiplied by the CT resolution (0.204 mm). ROI depth maps were created for each timepoint CT (n=4) for each mouse (n = 4).

### 3. *Image Deconvolution*

To reduce the blur intrinsic to coupled electron-optical photon transport in the mouse tissue, we performed iterative Lucy-Richardson deconvolution (10 iterations) with each of the CLI images with the a priori generated CSFs [23, 24]. The pixels within the CLI photo mask were given a weight of one in the Lucy-Richardson optimization while pixels outside the mask were provided a weight of zero. The weighting was performed to acknowledge that photons outside the boundary of the mouse would not follow the diffusion of the precomputed CSFs. While the mouse surface is not planar, for the purposes of the deconvolution, the CLI image was projected to a virtual imaging plane. This not true to the mouse geometry; however, it provides a first approximation that is useful in deblurring the luminescence for more realistic quantitative results.

A given CLI image was deconvolved with all CSFs generated at varying depths (2-10mm in 1mm increments) to create eight separate deconvolved images (**Figure 4d**). A weight for each image was computed by multiplying the total number of voxels In the ROI at the given depth by the normalized peak intensity of the CSF. A weighted average of the eight images was then computed to create the final deconvolved image as follows where $\circledast_{RL}$ indicates Richardson-Lucy deconvolution. Image weights were computed using voxel depth frequency in the ROI as well as CSF FWHM integral intensity.

$$CLI_{Deconvolved} = \sum_{n=3}^{10} w_n ( CLI \circledast_{RL} CSF_n) \qquad (7)$$

## 4. Activity Calculation

Deconvolution of the CLI images aided in restricting the Cherenkov signal to its planar origin. However, the deconvolution space is non-physical in contrast to our calibration images that were recorded in the original image space. To compute activity within an ROI, the deconvolved image was first masked using the ROI planar projection obtained previously from CT. We re-convolved the ROI-masked image with each CSF, using the same weights as the deconvolution process to compute a weighted average across the eight re-convolved images (**Figure 4d**).

Each pixel value in the ROI was divided by a depth-dependent calibration coefficient obtained from the calibration curve and previously computed ROI depth map to obtain an activity concentration representative of the ROI voxel stack projected onto the pixel. The activity concentration values were averaged and scaled by the ROI planar area to obtain ROI activity. Additionally, activity was calculated without the deconvolution step to assess the impact of the deconvolution-reconvolution scheme on estimated activity.

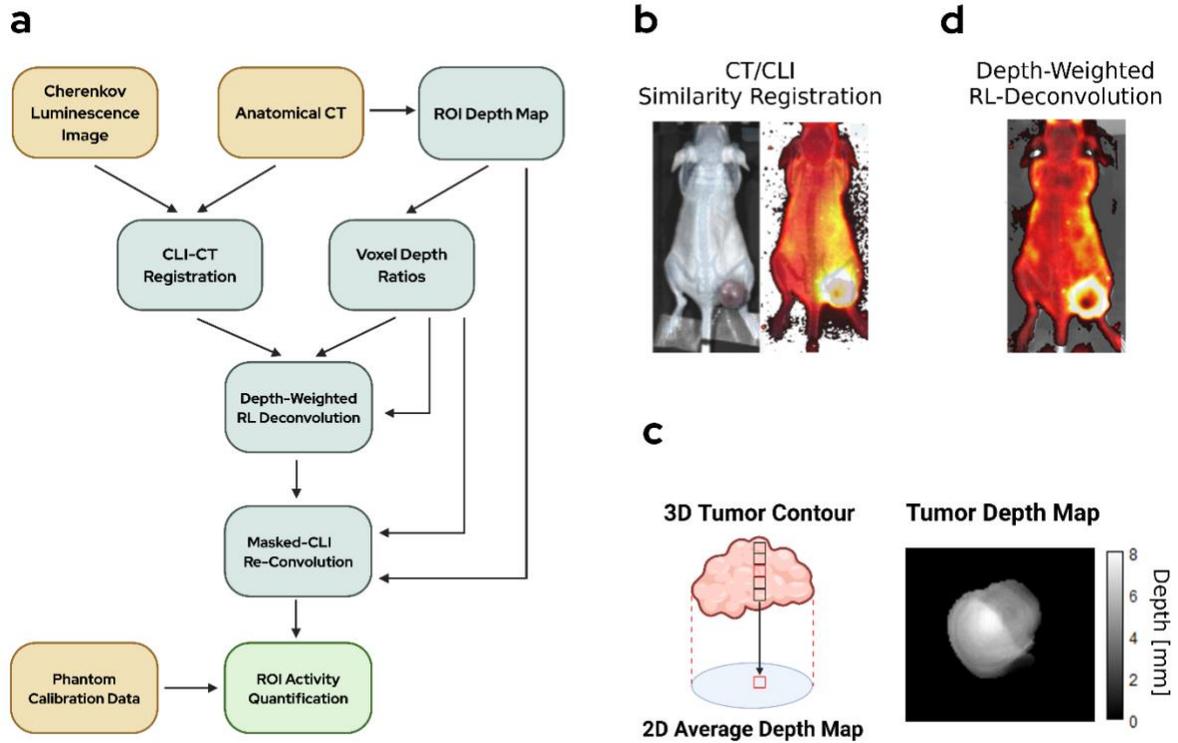

**Figure 4. (a) Workflow outline for Cherenkov-based activity quantification. (b) A similarity registration is performed between coronal masks of the CT and CLI images. (c) A planar projection of each ROI drawn on the CT is made where each pixel value represents the average depth (mm) below the surface of the mouse of the voxel stack projected to the pixel. (d) Depth-weighted RL-deconvolution is performed on the CLI images to limit signal blurring and crosstalk.**

**Dosimetry**

*PET-Based Dosimetry*

The Geant4-based, in-house MC platform *RAPID* was used to estimate dose distribution from the PET images [25]. Dose was calculated for $^{90}$Y, the therapeutic counterpart of $^{86}$Y, by applying a physical decay correction to account for the differing decay rates of the isotopes. CT images were used to define the voxelized geometry of the mouse while PET images were used to define the particle phase space. Absorbed dose in each voxel was calculated by integrating voxel-wise dose rates over time using the trapezoidal rule, assuming only physical decay after the final time point. Absorbed dose was averaged over each ROI to compute average dose to the liver and tumor.

*Cherenkov-Based Dosimetry*

Cherenkov-based dosimetry was also achieved using the *RAPID* platform. To compensate for the lack of a 3D activity distribution, we used the contoured CT and our in vivo estimates of ROI activity to synthesize Cherenkov-based PET scans. Three activity bearing compartments were defined: liver, tumor, and remainder. The tumor and liver compartments were uniformly filled with previously calculated ROI activities. For the remainder compartment, the decay-corrected Cherenkov radiance in the abdomen of the mouse was measured at each timepoint to estimate the biological clearance of $^{86}$Y-NM600. With the estimate of biological clearance, and the known physical decay rate, the total body activity over time was modeled using a bi-exponential equation. The measured liver and tumor activities were subtracted from the total to provide the activity in the remainder. The remainder activity was uniformly distributed throughout the mouse body.

# Results

## Phantom Study

A representative CL image of the well plate phantom is displayed in **Figure 5a.** Measurements of the phantom's Cherenkov radiance yielded quantitative data relating a source's depth in tissue to its corresponding observed surface radiance. Values for the analysis were averaged across triplicate plated wells with well standard deviation used to quantify uncertainty. Before analysis, we removed background by subtracting the radiance measured in the zero-activity wells. Background values remained relatively constant across all images (7902 $\pm$ 307 photons/s/cm$^2$/sr). **Figure 5b** shows the radiance [photons/s/cm$^2$/sr] generated in a well as a function of plated activity concentration [$\mu$Ci/cm$^2$] and depth [mm]. Linear regression of activity and observed radiance was performed at each intralipid-blood mixture depth **(Figure 5c)**. Across the surveyed range of 1.66-9.93 mm, we found the depth dependent calibration to follow a negative linear trend (Bias: 9734.2 Radiance/$\mu$Ci/cm$^2$; Slope; -613.8 Radiance/$\mu$Ci/cm$^2$/mm) with a Pearson correlation coefficient of r = 0.976 (**Figure 5d)**. In further analysis, the measured linear fit was implemented as a method to convert observed radiance from a source at a known depth to estimated activity.

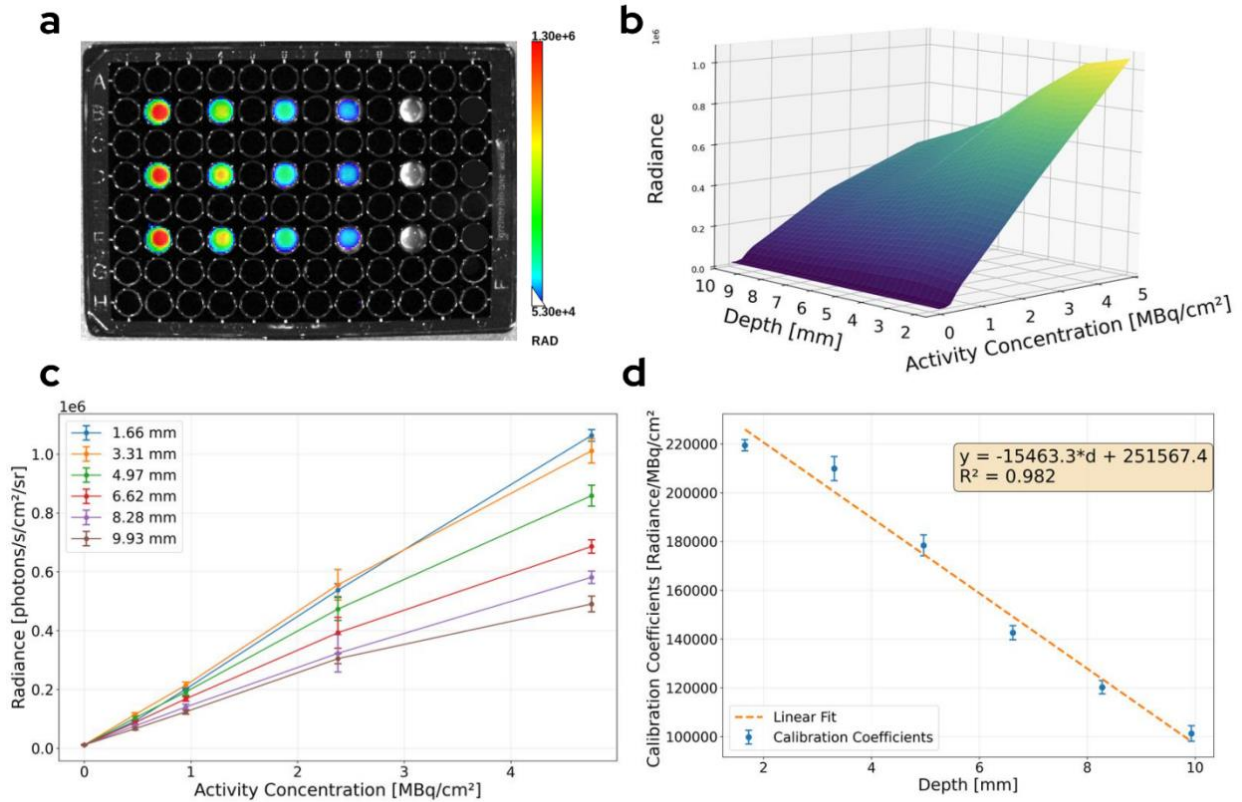

Figure 5. (a) displays a representative image of the calibration well plate with a depth of 3.31 mm intralipid-blood mixture added to each well. Activities from left to right are 0, 0.185, 0.370, 0.925, 1.850 MBq). A surface plot of the measured radiance across the varying activity wells (b) shows the general trend of increasing radiance as a function of activity and decreasing radiance as a function of depth. Isovolumetric curves (c) were graphed to obtain a linear calibration coefficient at each depth in tissue. A linear fit of the calibration coefficients was performed (d) to determine an expression for the calibration coefficient as a function of depth.

## GEANT4 Cherenkov Spread Functions

Cherenkov Spread Functions (CSF) describing the spread of Cherenkov light from a $^{86}Y$ positron source at varying depth in tissue (2-10 mm) were generated using GEANT4. **Figure 6a** shows the representation of the CSFs as planar slices of the Cherenkov photon propagation in tissue. The radial profiles of the generated CSFs are displayed in **Figure 6b.** The FWHM of the CSFs was observed to have a strong linear dependence on source depth increasing from 2.9 to 11.8 mm across source depths of 2-10 mm (**Figure 6c**). Integral intensity within the FWHM of the CSF decreased semi-linearly with depth, marked distinctively by an inflection point at 5 mm (**Figure 6d**). To reflect the higher radiance of shallow sources, deconvolution averages were weighted by the integral intensity and by the number of voxels at each depth. Cherenkov spread functions were normalized to unity integral intensity for use in deconvolution and re-convolution. Note that CSFs must be generated per isotope. While propagation of optical photons will remain the same, the characteristic beta spectrum of a given isotope will affect both total photons radiated as well as the spread due to beta particle transport.

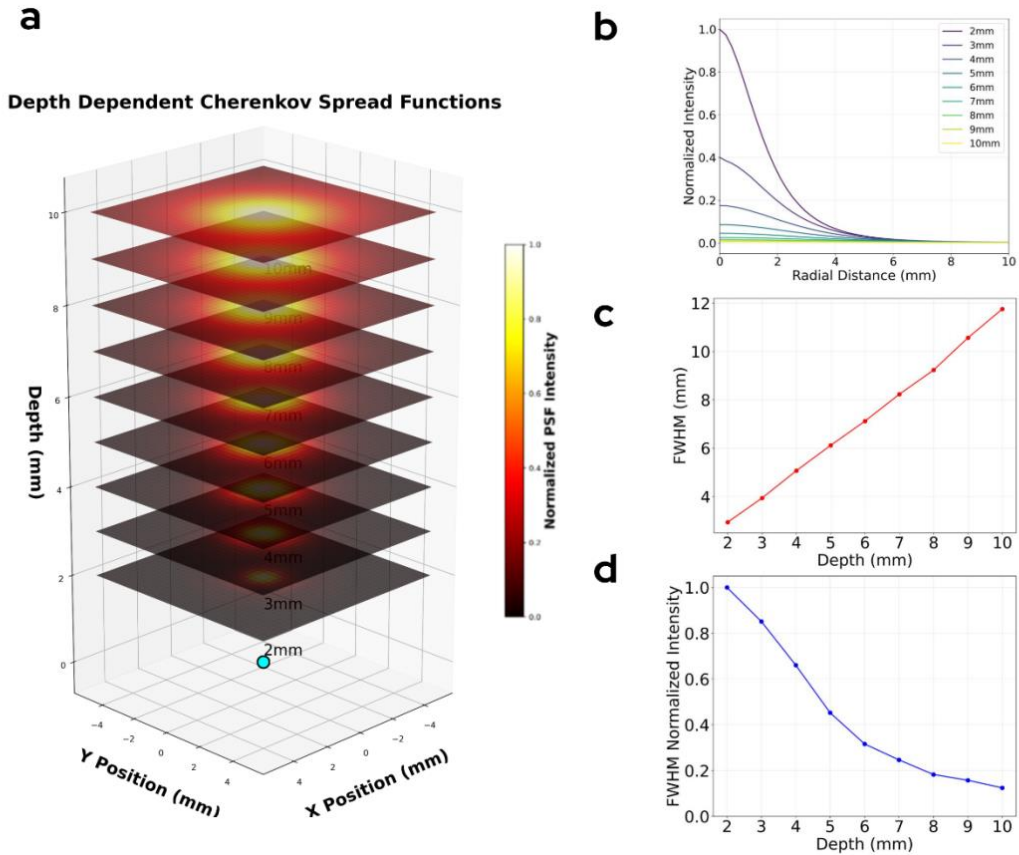

Figure 6. (a) shows the propagation of a photon through tissue forming the different Cherenkov Spread Functions (CSF) at each depth. Each CSF is normalized to its own maximum value. (b)The profile of each CSF normalized to the group maximum is shown. (c) The full width half maximum linearly of the CSF profile linearly increases with depth. (d) The integral intensity within the FWHM varies semi-linearly with a point of inflection at 5 mm depth in tissue.

**In Vivo Activity Estimation**

      Serial biodistribution of activity in the liver and tumor ROI was measured with both PET and CLI based methodologies at each imaging timepoint. For the CLI based measurements, activity was calculated two ways: using just the phantom based depth-based calibration and

using the deconvolution-reconvolution scheme in addition to depth dependent calibration. Representative PET and CLI images used to measure activity as well as activity estimation methodology comparisons are displayed in **Figure 7.** The average error at each timepoint between the CLI-based methodologies and PET quantification is displayed in **Table 2.** The agreement between the CLI and PET images is shown to increase when the deconvolution correction is applied. Average quantification error for the CLI deconvolution methodology across the first three timepoints was 15.4% and 10.3% percent for the liver and tumor, respectively. Relative error increases markedly at the fourth timepoint, however, absolute error is on the order of 5-10 kBq, highlighting the sensitivity limit of the CLI method due to mouse background auto luminescence.

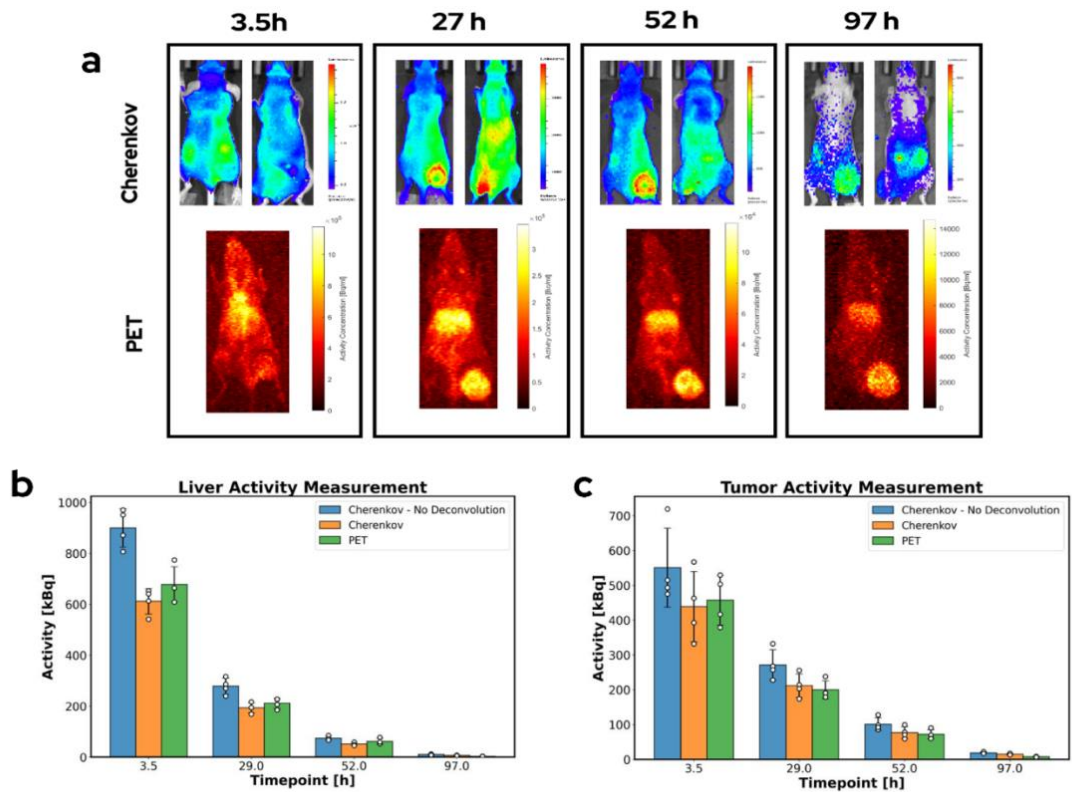

Figure 7. (a) CLI and PET images of a representative mouse show that signal is localized in the same regions in each modality. (b) Activity estimates made using both CLI and PET methodologies agree within one standard deviation. The CLI methodology including the deconvolution step shows better agreement with the PET images

Table 2. Percent error between CLI based activity quantification methodologies and PET ground truth across four biodistribution timepoints

| ROI | Timepoint [h] | CLI (No Deconv)/PET Error [%] | CLI (Deconv)/PET Error [%] |
|---|---|---|---|
| Liver | 3.5 | 33.7 | 12.7 |
|  | 29 | 33.1 | 15.9 |
|  | 52 | 26.7 | 17.5 |
|  | 97 | 255.8 | 125.2 |
| Tumor | 3.5 | 21.7 | 8.3 |
|  | 29 | 35.7 | 10.5 |
|  | 52 | 40.1 | 13.3 |
|  | 97 | 156.4 | 101.9 |

## Dose Estimation

Tumor dose was estimated by computing the absorbed dose rate at each timepoint using the validated Monte Carlo platform *RAPID*. Synthetic PET scans were created from the CLI activity estimates to calculate Cherenkov-derived dose. Physical decay was assumed at the last timepoint to carry the dose rate integration through the total isotope lifetime. All four timepoints were used for the PET dose calculation. However, only the first three timepoints were used to calculate CLI derived dose due to high relative error in the fourth timepoint. These errors are a result of low signal to background ratio after extended physical decay and biological clearance. Calculated dose rates and total absorbed doses are displayed in **Figure 8.** PET and Cherenkov based dose calculations yielded a population average of $3.2 \pm 0.2$ Gy/MBq and $3.4 \pm 0.3$ Gy/MBq, respectively. A paired T-test (n=4; df=3) yielded a P-value of 0.31 indicating that the two methodologies yielded dose results that cannot be stated to be significantly different.

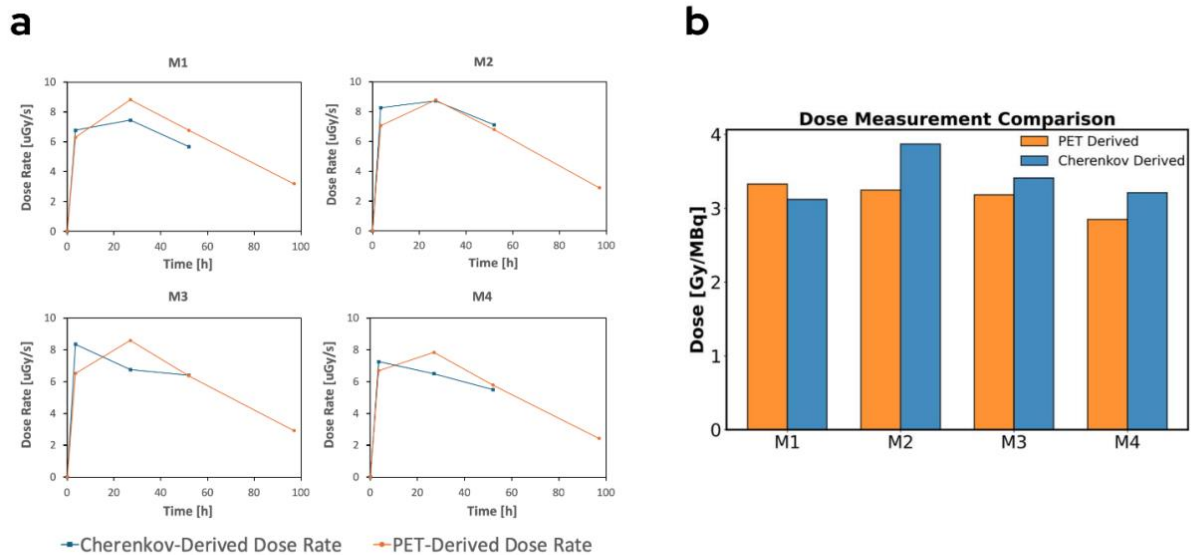

**Figure 8. (a) displays the monte carlo calculated dose rates at each timepoint for the PET and CLI derived synthetic PET scans (b) displays the integrated dose, assuming physical decay at the last timepoint.**

## Discussion and Conclusions

Although in vivo CLI imaging was first reported nearly 15 years ago, it remains a largely underutilized modality for preclinical dosimetry [6]. The success of novel radiopharmaceutical agents has led to a surge in preclinical RPT studies, creating a need for equally innovative dosimetry approaches. While CLI cannot yet match PET/SPECT in terms of high-resolution, tomographic biodistribution imaging, it can complement conventional nuclear imaging techniques and support expansive pre-clinical RPT trials.

PET- and SPECT-based murine dosimetry typically relies on small "surrogate" cohorts to derive population-level dose statistics; the resulting prescriptions are then applied to separate efficacy cohorts. Without retrospective dosimetry, however, dose–response relationships remain uncertain, as pharmacokinetics and binding affinity can vary across—or even within—subjects during therapy. Scaling PET/SPECT to large cohorts is impractical: scanning four mice takes 20–60 min, depending on total-body activity. CLI, however, is a cheap and high-throughput alternative to PET and SPECT enabling either more biodistribution timepoints to be recorded or dosimetry to be performed for much larger cohorts of mice than is possible in the standard small animal imaging facility. The methodology proposed in this paper can enable high throughput dosimetry.

Prior studies have made progress towards quantitative CLI through depth-dependent attenuation corrections as well as relative measurements relating in vivo radiance to a final timepoint ex vivo measured activity [14, 26]. While these methods account for the decrease in intensity due to high scattering optical photon transport, they do not account for conflation of nearby signals due to the diffusion of Cherenkov photons into neighboring regions. When only accounting for the attenuation of Cherenkov radiance, organs located close to one another will be overcounted resulting in higher activity and dose estimates. This error is time-dependent and cannot be accounted for by a uniform correction factor.

Building on earlier work, we address photon diffusion with an iterative deconvolution that uses Monte-Carlo-generated Cherenkov spread functions, coupled to a model-agnostic full-spectrum attenuation correction. The use of $^{86}$Y-NM600 provided an optimal test case due to its dual CLI and PET compatibility. High-energy positrons emitted by $^{86}$Y produce substantial Cherenkov signals while simultaneously allowing PET visualization and quantification. The correlation between CLI and PET-derived activities in liver and tumor ROIs supports the validity of our calibration and deconvolution scheme. Notably, the use of deconvolution reduced quantification error by approximately 50% in both tissue compartments across the first three timepoints. While signal-to-noise limitations at later timepoints degraded performance, this is expected due to both radiotracer clearance and increasing influence of background auto-luminescence.

Our dosimetric comparison further substantiates the viability of quantitative CLI. Tumor-absorbed dose estimates from CLI-derived synthetic PET images yielded values statistically indistinguishable from those derived from true PET images, with a mean population difference of only ~6%. These results are particularly encouraging given the planar nature of CLI and the fact that only three timepoints were used for CLI dose estimation versus four for PET.

Certain limitations of our CLI method merit discussion. First, as CLI radiance is most prominent for sources near the surface, our activity quantification workflow may not be as applicable to small, deep seated organs and tumors. Second, our CLI-based reconstruction assumes a homogeneous optical model within each ROI, which may not capture intra-organ heterogeneity in optical properties. Third, the spatial resolution of CLI is fundamentally limited by both the diffusion of optical photons and the resolution of the imaging system. Although our deconvolution method improves spatial localization, fine-scale features will still be blurred limiting dosimetry to the organ level. Fourth, the need for anatomical CTs images limits throughput, however, in future studies it is likely that only a single CT is needed instead of for each timepoint. As only local registration of ROIs is necessary for activity estimation, similarity transforms added by manual alignment of ROIs should provide the anatomical reference needed for non-CT matched CLI timepoints. Additionally, reasoned estimations of tumor depth and planar area could be used in place of CT based information.

Looking ahead, this work lays the groundwork for high-throughput preclinical dosimetry using CLI. Future extensions of this work will test the quantitative accuracy of our approach on therapeutically relevant isotopes such as $^{90}$Y and $^{177}$Lu. Additionally, we aim to expand our methodology to Cherenkov producing decay chains from alpha emitters such as $^{225}$Ac and $^{212}$Pb to provide individualized preclinical alpha emitter tumor dosimetry.

In conclusion, our results demonstrate that with appropriate correction for tissue attenuation and photon scatter, CLI can yield quantitative biodistribution and dosimetric information comparable to PET. This approach significantly expands the scalability and accessibility of in vivo dosimetry, providing a valuable tool for accelerating preclinical RPT development.

# Acknowledgements

Figures 1-8 created in Biorender. Haasch, C. (2025) https://BioRender.com/8c3g765

The authors would like to acknowledge the Cancer Center Support Grant: NCI P30 CA014520, University of Wisconsin Small Animal Imaging & Radiotherapy Facility and NIH S10OD028670-01 for supporting this work."

# Statements & Declarations

**Funding**

This work was supported by NIH 5P01CA250972-05.

**Competing Interests**

The authors declare that they have no competing interests.

**Author Contributions**

Campbell Haasch, Sydney Jupitz, and Bryan Bednarz contributed to the study conception and design. Reinier Hernandez and Brian Pogue contributed to study design and methodology development. Material preparation, data collection, and analysis was performed by Campbell Haasch. Aubrey Parks and Malick Bio Idrissou contributed to methodology development. The first draft of the manuscript was written by Campbell Haasch. All authors read and approved the final manuscript.

**Ethics Approval**

Approval for animal studies was granted by the University of Wisconsin-Madison School of Public health and Medicine IACUC on September 18, 2023.